\documentclass[onetwocolumn,showpacs,preprintnumbers,amsmath,amssymb]{revtex4}
\usepackage{graphicx}
\usepackage{bm}

\newcommand{\ie}{{\it i.e.}}
\newcommand{\eg}{{\it e.g.}}

\def\bra#1{ \langle #1\!\mid }
\def\ket#1{\mid\!#1\rangle}

\def\ave#1{\langle~#1~\rangle}

\begin{document}

\title{The role of coherence on two-particle quantum walks}

\author{Li-Hua Lu$^1$, Shan Zhu and You-Quan Li$^{1,2}$}
\affiliation{1. Zhejiang Institute of Modern Physics and Department of Physics,\\ Zhejiang
University, Hangzhou 310027, P. R. China\\
 2. Collaborative Innovation Center of Advanced Microstructures, Nanjing, P. R. China
}

\begin{abstract}
  We investigate the dynamical properties of the two-bosons quantum walk in system with  different degrees of coherence, where  the effect of the coherence on the two-bosons quantum walk  can  be naturally introduced.   A general analytical expression of the two-bosons correlation function for both pure states and mixed states is given. We propose a possible two-photon quantum-walk scheme with a mixed initial state and find that the two-photon  correlation function and the average distance between two photons can be influenced  by either the initial photon  distribution, or the relative phase, or the degree of  coherence.  The propagation features  of our numerical results  can be explained by our analytical two-photon correlation function.

\end{abstract}
\received{\today}
\maketitle

\section{Introduction}
As the quantum mechanical counterparts of the classical random walk~\cite{Ahar},
the quantum random walk has been increasingly receiving  attentions
because of their potential applications range from   quantum information to simulation of physical phenomena. For example, the quantum walk offers an advanced tool for building quantum algorithm~\cite{Moh,Shen,Sal}
that is shown to be of the primitive for universal quantum computations~\cite{und,Lov,chi,and}.
We know that  the quantum walk include two main classes that are discrete-time quantum walk and continuous-time quantum walk~\cite{Jwa,Far}.
The continuous-time quantum walk can evolve continuously with time through tunneling between neighbors sites and does not require quantum coin to generate superposition of states.
This means the continuous-time quantum walk can be implemented
via a constant tunneling of quantum particles in several possible lattice sites.
So far,  the quantum walks of single particles have been studied in experiments by using either classical waves~\cite{peret}, single photons~\cite{Sch,broo}, or single atoms~\cite{Kars,Wei}.
Additionally,  quantum walks of two correlated photons trapped in waveguide lattices
were also  studied in experiments~\cite{Yar,Alberto}.

Note that  many-particle quantum walks can exhibit more fascinating quantum features in contrast to single-particle quantum walks.  The reason is that single-particle quantum walks can be exactly  mapped to classical wave phenomena~\cite{Knight} but for quantum walks of more than one  indistinguishable particle, the classical theory can not provide sufficient descriptions. In Ref.~\cite{omar} , the authors theoretically studied the discrete-time quantum walk of two particles and demonstrated the distinctly nonclassical correlations. Meanwhile,  the effect of interactions between particles on quantum walk of two indistinguishable particles was theoretically studied in Ref.~\cite{lahini,Qin}, where the system was assumed  to be completely coherent and the influence of the degree of coherence was not studied. We know that except for the interaction between particles, the other factors, \eg, the initial states, the quantum-walk parameters and the degree of coherence  of the system,  can also affect the features of two-particle quantum walks.  Especially, we know that the major challenge to experimentally realize quantum walks of correlated particles is to find a low-decoherence system that can preserves the nonclassical features of quantum walks~\cite{Alberto}, which implies that the influence of the degree of  coherence of the system on  two-particle quantum walks is important.  Then it is worthwhile to investigate the properties of two-particle quantum walks with attention to different  degrees of coherence since the decoherence effects in quantum walks have potential algorithmic applications~\cite{kendon}.

In this paper, we propose  a density matrix formulism  to study the two-particle quantum walk
where the degrees of coherence can be naturally introduced.
With the help of Heisenberg equation of motion, we derive a general analytical expression
of the two-particle correlation function.
As a concrete example, we propose a possible two-photon quantum-walk scheme with a mixed initial state to exhibit the quantum features  of the two-particle  quantum  walk
via the two-particle  correlation and the average distance between the two particles.
Our result exhibits that the propagation of the two particles depends
not only on the initial distribution of the two particles
but also on the relative phase and  the degree of  coherence of the system.
Such a propagation  feature  can be
explained  by our analytical two-particle correlation function.  In the next section, we present  the model and derive the  analytical expression of the two-particle correlation. In Sec.~\ref{sec:two-photon}, we propose a concrete scheme  to show some dynamical features of two-particle quantum walks. Our main conclusions are summarized in Sec.~\ref{sec:conc}.

\section{A general formulation }
We consider a two-particle quantum walk in a one-dimensional lattice space.
The propagation of the two particles is described by the evolution of the state
of a tight-binding model,
\begin{equation}
H=-\sum_{q}T_{q,q+1}(\hat{a}_q^\dagger\hat{a}_{q+1} + \mathrm{h.c.})
+\sum_q \beta_q\hat{a}_q^\dagger\hat{a}_q,
\end{equation}
where the operators $a_{q}^\dag$ and $a_{q}$  create and  annihilate a bosonic  particle at site $q$, respectively.
Here the parameter $T_{q,q+1}$ refers to the tunneling strength of particles  between the nearest neighbor sites, and
\begin{equation}
\beta_q=T_{q+1,q}+T_{q-1,q}.
\end{equation}
 Note that the above form of  $\beta_q$ was  picked  to keep the probability conservation in the proposal of continuous-time quantum walk via decision tree~\cite{Far} . Now more generally, the value of $\beta_q$ can be arbitrary due to the  probability conservation is naturally satisfied in quantum mechanics.
If the tunneling strength $T_{q,q+1}$ is a constant, $\beta_q$ will become  a constant for the periodical boundary condition. In this case, the value of $\beta_q$ does not affect the dynamical properties of the quantum-walk system. Whereas, for the open boundary  condition, the values of $\beta_q$ for the two boundary sites are different from that for the other sites. This can naturally introduce  two defects to the quantum-walk  system. Note that the effect of defects on single-particle quantum walks was studied in Ref.~\cite{Li}.

Since we consider a two-bosons  system,
the Fock bases describing the system are
\begin{equation}
\label{eq:fockstate}
\ket{1}_q\ket{1}_r=\frac{1}{\sqrt{1+\delta_{q,r}}}\hat{a}_q^\dagger\hat{a}_r^\dagger\ket{\mathrm{vac}},
\end{equation}
where $\delta_{q,r}$ denotes the Kronecker delta.
Equation (\ref{eq:fockstate}) represents a two-particle state with one
on the $q$th site and the other one on the $r$th site.
Note that $\ket{1}_q\ket{1}_r$ is regarded as identical to $\ket{1}_r\ket{1}_q$
for indistinguishable particles that we considered.
Meanwhile, the two particles can be in the same site (\ie, $q=r$)
for bosonic particle that we considered.
Thus the Hilbert  space expanded by the aforementioned Fock bases is of $D=L(L+1)/2$ dimension.
Here $L$ denotes the number of the sites.
We know that the propagation of the two particles is determined not only by the property of the waveguide lattice but also by the two-particle input state.
If the two particles are in a pure state at the initial time,
the two-particle input state can be expressed as a wavefunction, namely, a coherent superposition of the Fock bases,
\begin{equation}\label{eq:purstates}
\ket{\psi}=\sum_{q,r}c_{q,r}\ket{1}_q\ket{1}_r,
\end{equation}
where  $\sum_{q,r}|c_{q,r}|^2=1$.
However, if the two particles are in a mixed state at the initial time,
the two-particle input state needs to be described by a density matrix
rather than wavefunction.
Such a density matrix is given by
\begin{equation}
 \label{eq:densitymatix}
 \rho=\sum_{qr,q'r'}\rho_{qr,q'r'}\bigl(\ket{1}_q\ket{1}_r\bigr)\bigl(\bra{1}_{q'}\bra{1}_{r'}\bigr),
\end{equation}
which is a $D\times D$ matrix.
We know that $\textrm{Tr}\rho^2\leq(\textrm{Tr}\rho)^2$
where the equal sign holds only for pure states.
In the following,
we will focus on the two-particle quantum walk for the mixed input states.

Now we are in the position to study the propagation of the two particles
with the help of Heisenberg  equation of motion for the creation operators,
namely,
\begin{equation}
\label{eq:dye}
i\frac{\partial \hat{a}_q^\dagger}{\partial t}=\beta_q\hat{a}_q^\dagger+T_{q,q+1}\hat{a}_{q+1}^\dagger
+T_{q,q-1}\hat{a}_{q-1}^\dagger,
\end{equation}
where we set $\hbar=1$ for simplicity in calculation.
The creation operator $\hat{a}_q^\dagger$ at any time can be obtained with the help of Eq.~(\ref{eq:dye}),
\begin{equation}
\label{eq:creation}
\hat{a}_q^\dagger(t)=\sum_r U_{q,r}(t)\hat{a}_r^\dagger(0),\quad
U(t)=e^{-iHt},
\end{equation}
where  $U_{q,r}(t)$ is the probability amplitude of a single particle transiting
from the $q$th waveguide to the $r$th one.
To exhibit the quantum behaviors of the two-particle quantum walk,
let us firstly evaluate the two-particle correlation function $\Gamma_{k,l}(t)=\ave{\hat{a}_k^\dagger(t)\hat{a}_l^\dagger(t)\hat{a}_l(t)\hat{a}_k(t)}$
which manifests the probability that the  two particles are coincident in the $k$th and the $l$th waveguide~\cite{Yar,Mattle}.
Since the two-particle input state can be described by the density matrix given
in Eq.~(\ref{eq:densitymatix}), the expectation value of any observable of the system
can be calculated via $\ave{\hat{O}(t)}= \textrm{Tr}(\hat{O}(t)\rho)$.
Then we obtain an expression of
two-particle correlation function
\begin{widetext}
\begin{eqnarray}\label{eq:twocorre}
&&\Gamma_{k,l}(t)=\sum_{q\neq r,q'\neq r'}\rho_{qr,q'r'}\Bigl(U_{kq'}U_{lr'}U_{lq}^*U_{kr}^*+U_{kq'}U_{lr'}U_{lr}^*U_{kq}^*
+U_{kr'}U_{lq'}U_{lq}^*U_{kr}^*+U_{kr'}U_{lq'}U_{lr}^*U_{kq}\Bigr)
  \nonumber\\
&&+\sum_{q,q'\neq r'}\sqrt{2}\rho_{qq,q'r'}\Bigl(U_{kq'}U_{lr'}U_{lq}^*U_{kq}^*+U_{kr'}U_{lq'}U_{lq}^*U_{kq}^*\Bigr)
+\sum_{q\neq r,q'}\sqrt{2}\rho_{qr,q'q'}\Bigl(U_{kq'}U_{lq'}U_{lq}^*U_{kr}^*+U_{kq'}U_{lq'}U_{lr}^*U_{kq}^*\Bigr)
\nonumber\\
&&+\sum_{q,q'}2\rho_{qq,q'q'}U_{kq'}U_{lq'}U_{lq}^*U_{kq}^*,
\end{eqnarray}
\end{widetext}
which presents a general form for either pure initial input states or mixed ones.
One can obtain the two-particle correlation at any time
as long as the density matrix corresponding to the input state is given. with the help of such an expression of two-particle correlation function, many dynamical features of two-particle quantum walk can be explained.

\section{Two-photon quantum walk for a concrete  input  state}\label{sec:two-photon}

In order to expose the quantum properties of two-particle quantum walks more clearly, we turn to a concrete example where the particles are assumed to be photons.   We know that each  beam can become two coherent beams after propagating through a grating~\cite{sza} and the  pure two-photon input states can be experimentally realized via injecting two coherent beams into waveguide lattice~\cite{Yar}.  Then we suppose that there are two incoherent  light beams  and the relation of their intensity is $\cos^2{\delta}$:$\sin^2{\delta}$. The two incoherent  beams propagate through two gratings, respectively, and then  simultaneously inject into the waveguide arrays.  The two incoherent beams  will create two pure two-photon states $\psi_1$ and $\psi_2$, respectively~\cite{Yar,sza}.   Because the initial two beams are not coherent, the initial state of the system needs to be  described by the density matrix
\begin{equation}
\rho=\cos^2\delta \ket{\psi_1}\bra{\psi_1}
   +\sin^2\delta  \ket{\psi_2}\bra{\psi_2}.
\end{equation}

As an example, we take $\psi_1=\cos{\frac{\theta}{2}}\ket{2}_1+\sin{\frac{\theta}{2}}e^{i\phi}\ket{2}_0$
and $\psi_2=\ket{2}_1$, then the initial density matrix is
\begin{align}
\label{eq:exrho}
\rho = \rho_{00,11}\ket{2}_0\bra{2}_1 + \rho_{11,00}\ket{2}_1\bra{2}_0
        + \rho_{11,11}\ket{2}_1\bra{2}_1 + \rho_{00,00}\ket{2}_0\bra{2}_0
\end{align}
 with $\rho_{00,11}=\cos^2\delta\cos{\frac{\theta}{2}}\sin{\frac{\theta}{2}}e^{i\phi}$,
$\rho_{11,00}=\rho_{00,11}^*$,
$\rho_{11,11}=\cos^2\delta\cos^2{\frac{\theta}{2}}+\sin^2\delta$,
and
$\rho_{00,00}=\cos^2\delta\sin^2{\frac{\theta}{2}}$.
Here $\ket{2}_q$ stands for
$\ket{1}_q\ket{1}_q$ whose definition has been given
in Eq.~(\ref{eq:fockstate}).   We can find that the  other matrix elements of $\rho$
are zeros except for the above four elements. Since the above four elements can be changed via $\delta$, $\theta$ and $\phi$,
without losing the generality, we redefine  them as
$\rho_{00,00}=\alpha$, $\rho_{11,11}=1-\alpha$ and  $\rho_{00,11}=\rho_{11,00}^*=e^{i\phi}\sqrt{\eta-1-4\alpha^2+4\alpha}/2$ with $0\leq\alpha\leq 1$. Here the parameter $\eta=2\textrm{Tr}(\rho^2)-1$ $(0\leq\eta\leq1)$ is introduced to characterize the degree of coherence~\cite{lhlu}.
We have $\eta=1$ when the system is in a pure state,  otherwise $\eta<1$.  With the help of Eq.~(\ref{eq:twocorre}),
the two-photon correlation of the system for the mixed state shown in Eq.~(\ref{eq:exrho}) yields
\begin{align}
\label{eq:extwocre}
\Gamma_{q,r} =  2\gamma\textrm{Re}(e^{i\phi}U_{q0}U_{r0}U^*_{r1}U^*_{q1})
         +2\alpha |U_{r0}U_{q0}|^2+2(1-\alpha)|U_{r1}U_{q1}|^2,
\end{align}
where $\gamma=\sqrt{\eta-1+4\alpha(1-\alpha)}$.
This implies that the two-photon correlation depends not only on the initial probability distribution of the two photons but also on the degree of coherence and the relative phase of the system at the initial time. The first term in Eq.~(\ref{eq:extwocre}) is a coherent one that  reveals well  the quantum nature of two-photon quantum  walk.  Taking a pure initial state (\ie, $\eta=1$) as an example,  the two-photon correlation function~(\ref{eq:extwocre}) becomes  $\Gamma=|\sqrt{2\alpha}e^{i\phi}U_{r0}U_{q0}+\sqrt{2(1-\alpha)}U_{r1}U_{q1}|^2$, which implies that the two-photon correlation can take place when the two photons  from the 0th site propagate to the $q$th and the $r$th sites, respectively,  or when the two photons from  the 1th site to the $q$th and the $r$th sites, respectively. Due to the two photons are indistinguishable, the two paths can interfere, which  is essentially the Hanbury Brown Twiss(HBT) interference~\cite{Han}.

Now we investigate the quantum features of a two-photon quantum walk
by considering a waveguide arrays consisting of (2$l$+1) identical waveguides.
In this case, the tunneling strengths between nearest-neighbor arrays are all the same, \ie, $T_{q,r}=C$ with $C$ being a constant, and $\beta_q$ becomes a constant $2C$ for the periodical boundary condition we considered.
Then $U_{q,r}(t)$ becomes $e^{i2Ct}i^{q-r}J_{q-r}(2Ct)$
where $J_q$ is the $q$th order Bessel function~\cite{Led,Yariv}.
With the help of Eq.~(\ref{eq:extwocre}),
we can write out the two-photon correlation function in terms of Bessel functions.
\begin{widetext}
\begin{align}
\label{eq:twocrebf}
\Gamma_{q,r}(\tau) =-2\gamma\cos{\phi} J_q(\tau)J_r(\tau)J_{r-1}(\tau)J_{q-1}(\tau)
 +2\alpha [ J_q(\tau) J_r(\tau) ]^2
   +2(1-\alpha)[ J_{r-1}(\tau) J_{q-1}(\tau) ]^2.
\end{align}
\end{widetext}
where $\tau=2Ct$.
\begin{figure}[tbph]
\includegraphics[width=48mm]{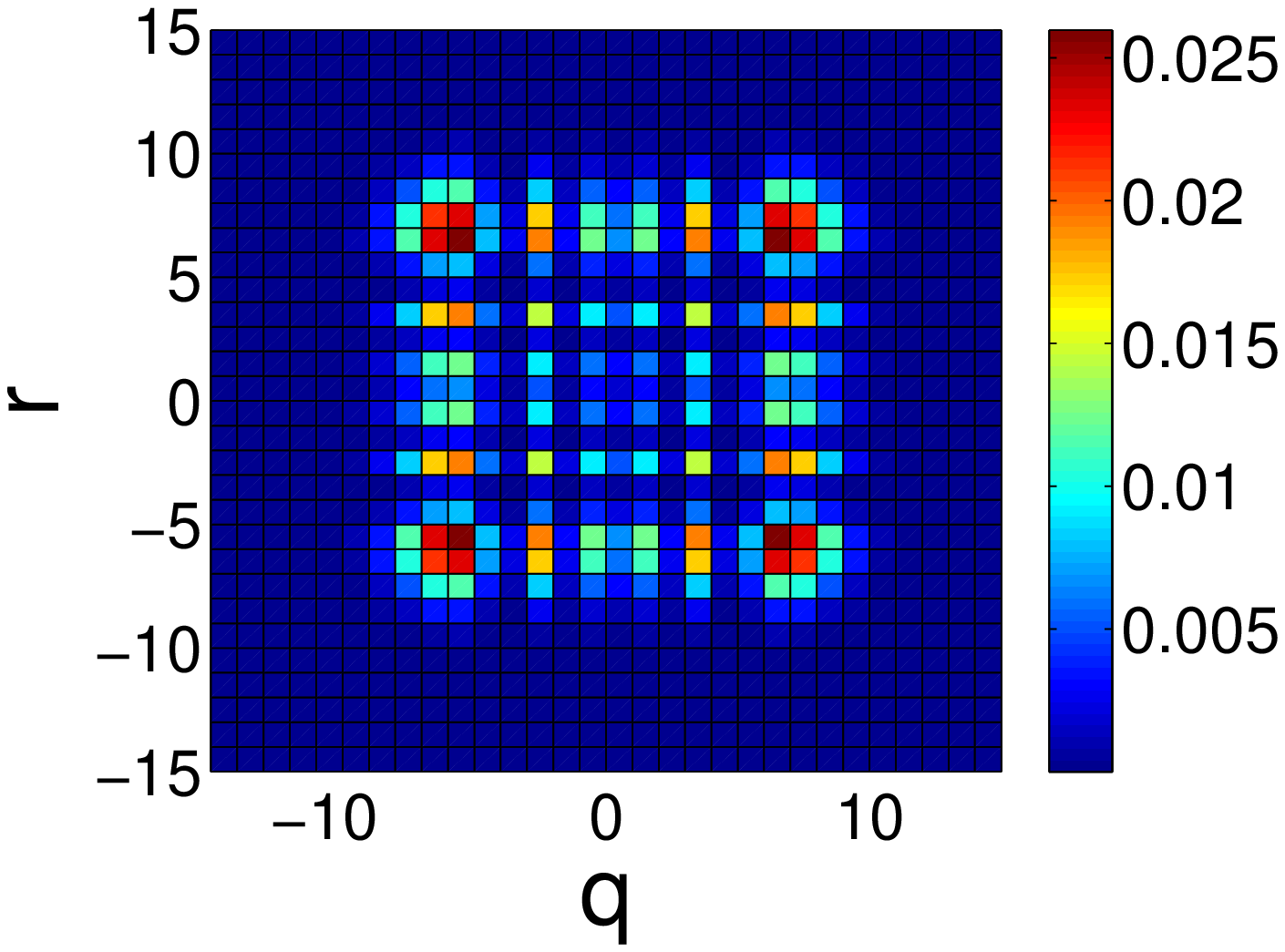}(a)
\includegraphics[width=48mm]{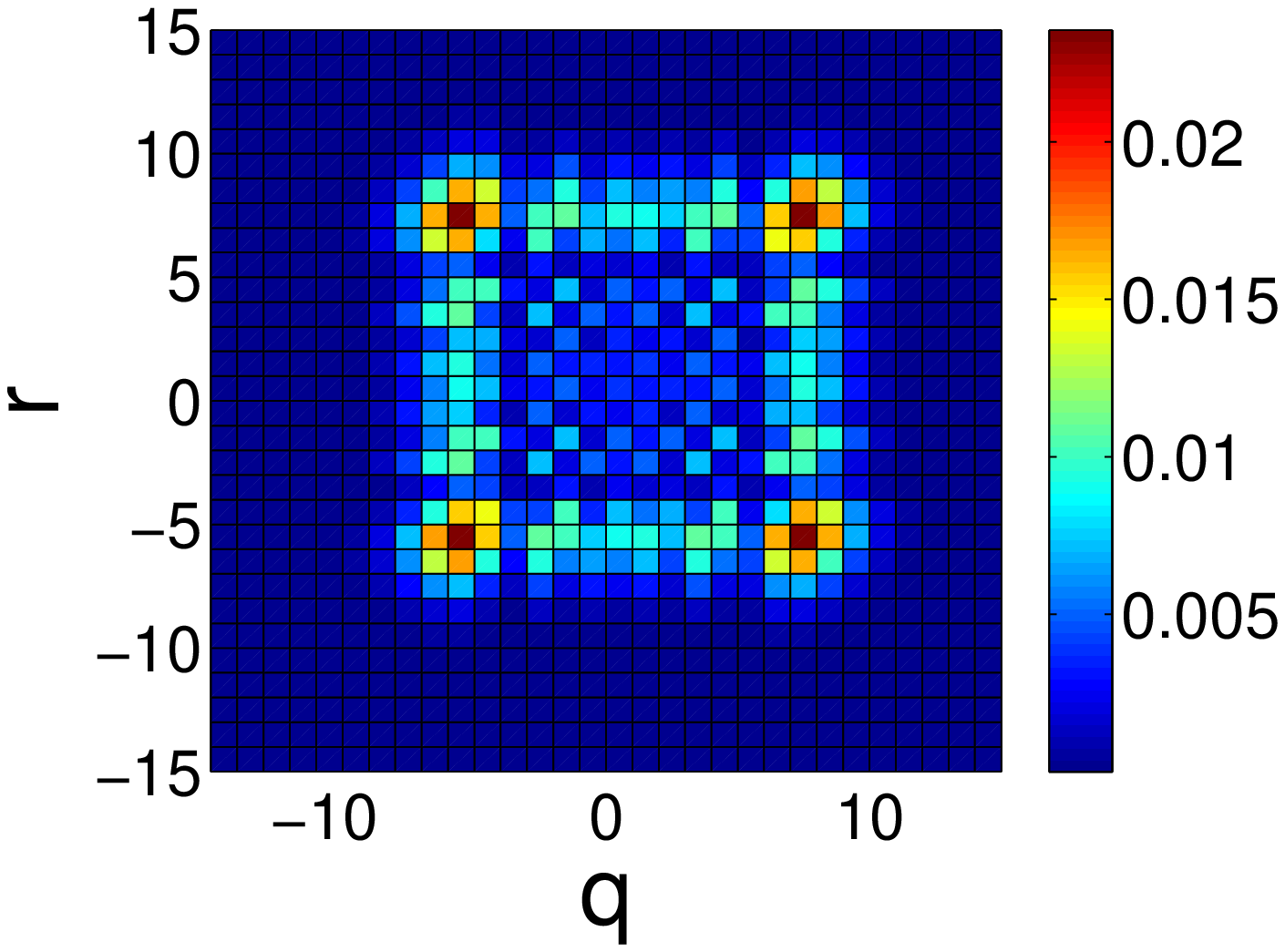}(b)
\includegraphics[width=48mm]{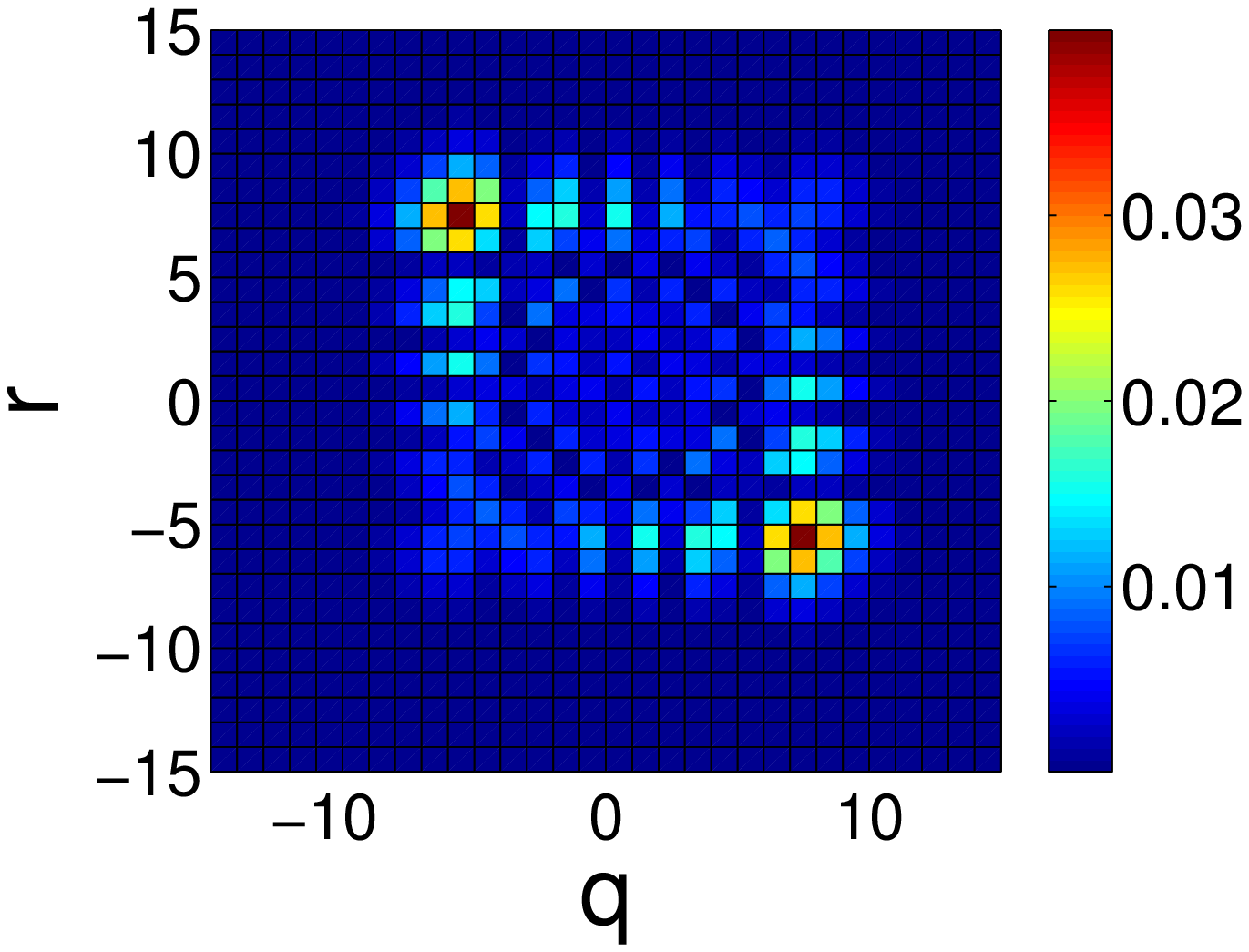}(c)
\includegraphics[width=48mm]{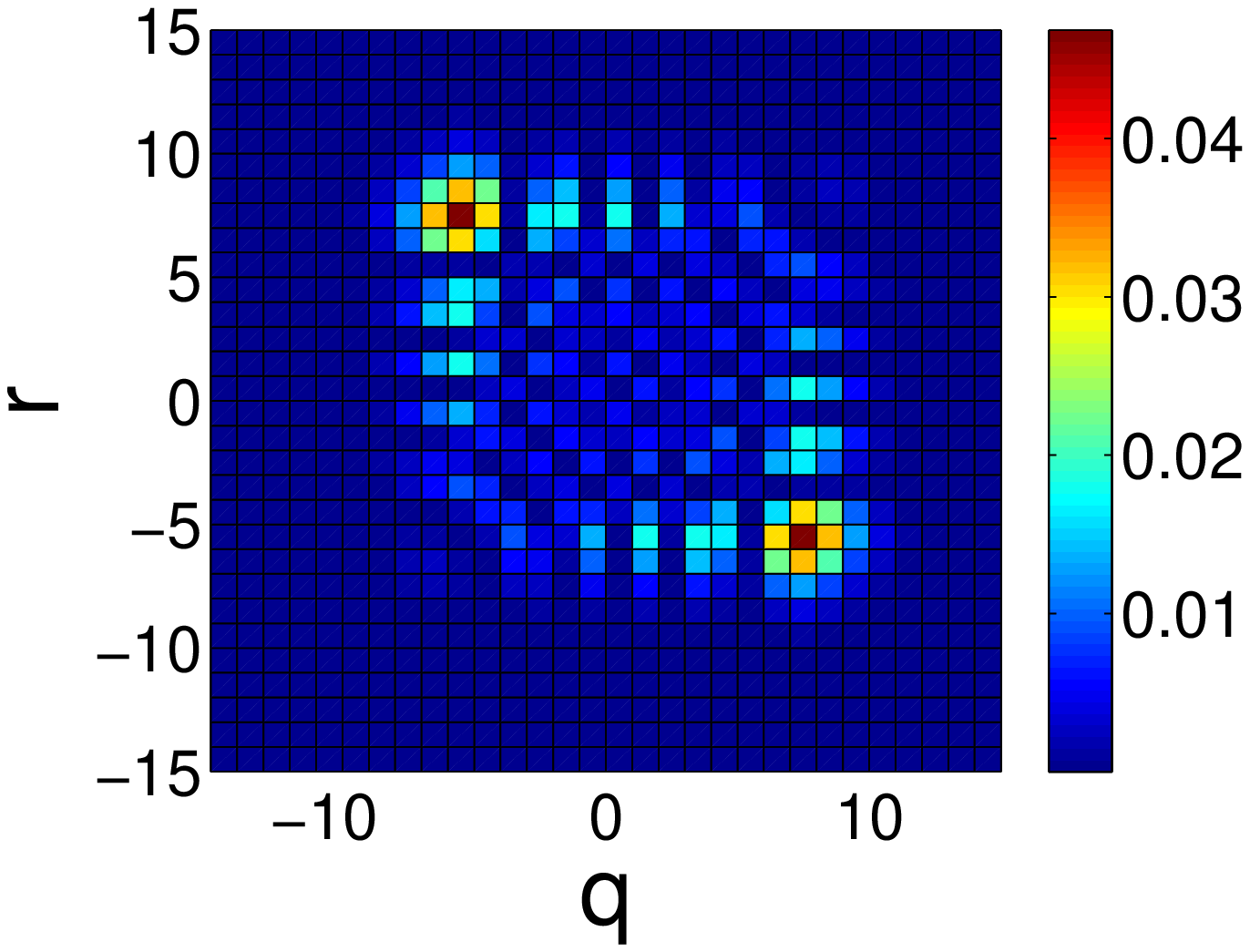}(d)
\includegraphics[width=48mm]{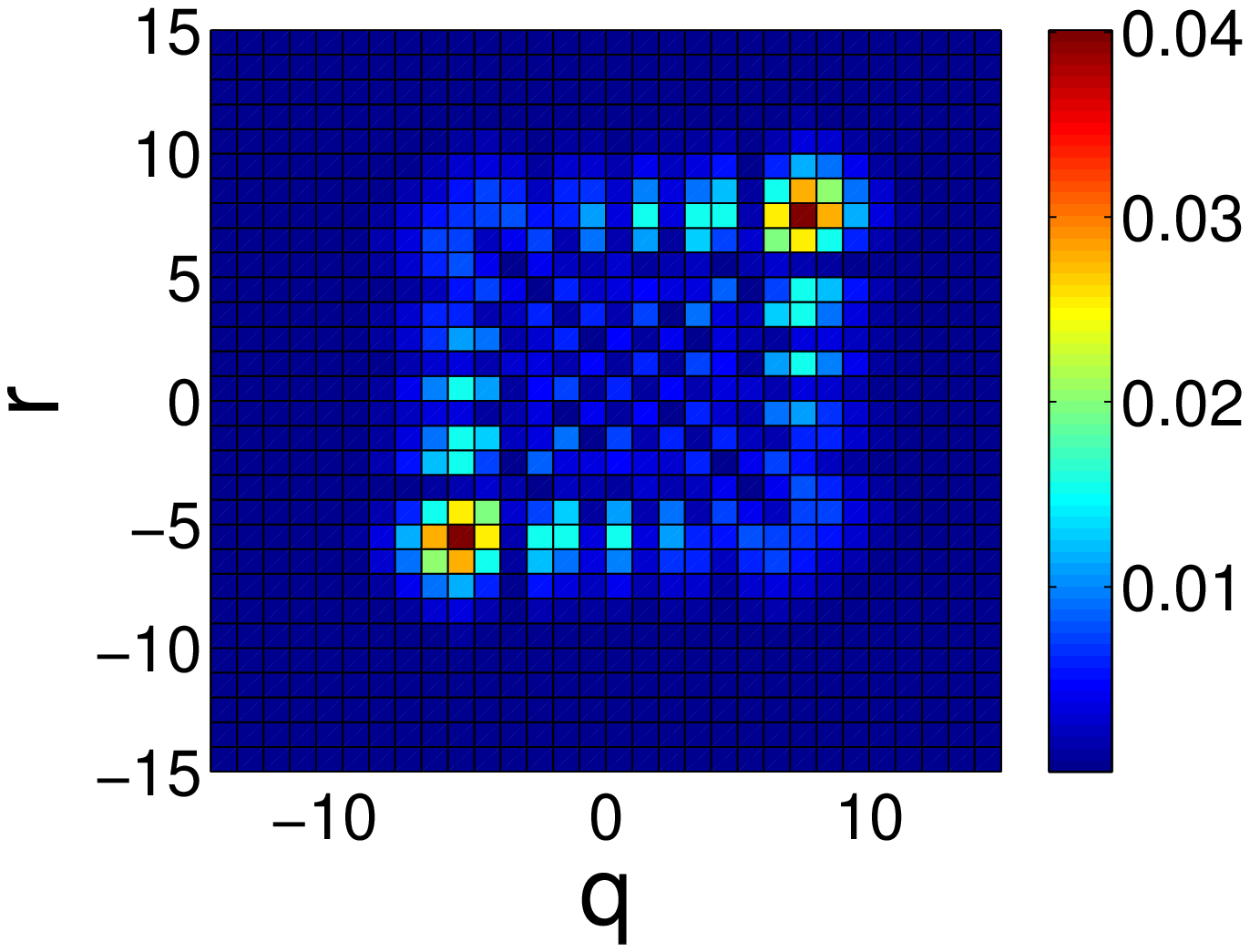}(e)
\includegraphics[width=48mm]{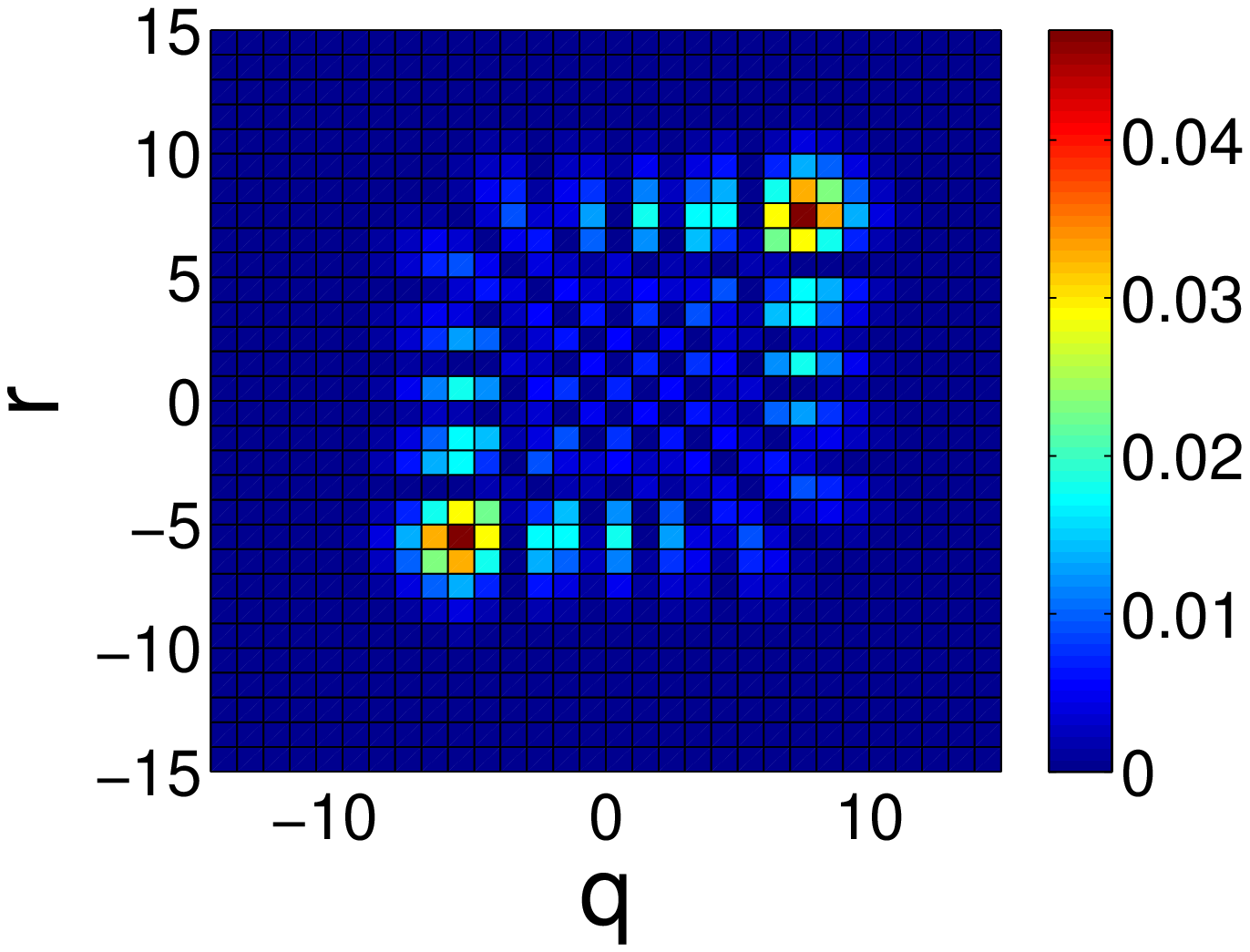}(f)
\caption{(Color online) Two-photon  correlation at time $t=4(1/C)$ for different initial conditions.  The initial condition is (a) $\alpha=1$, $\eta=1$,  $\phi=0$ (b) $\alpha=0.5$,  $\eta=0$, $\phi=0$
(c) $\alpha=0.5$,  $\eta=0.5$, $\phi=0$ (d) $\alpha=0.5$,  $\eta=1$, $\phi=0$ (e) $\alpha=0.5$,  $\eta=0.5$, $\phi=\pi$,   and  (f) $\alpha=0.5$,  $\eta=1$, $\phi=\pi$.  }
\label{fig:tcou}
\end{figure}

In Fig.~\ref{fig:tcou},
we plotted the two-photon correlation matrix at  time $t=4({1}/{C})$ for different initial conditions,  where $1/C$ is the unit of time (\ie, the time $t$ is in the unit of the inverse of tunneling strength).  We can see that each particle can be found on either side of origin after propagation, which is reflected in the four symmetric peaks in Fig.~\ref{fig:tcou} (a). For this case (\ie, $\alpha=1$  and $\eta=1$), the system is in  a pure state and Eq.~(\ref{eq:extwocre}) becomes $\Gamma_{q,r}=2|U_{r0}U_{q0}|^2$ which is the same as the result in Ref.~\cite{Yar}. Such a  correlation function is just a product of the two classical probability distribution, so there is no interference and the photons propagate in the ballistic direction. From Fig.~\ref{fig:tcou} (b), we can find that just like Fig.~\ref{fig:tcou} (a), the two photons also  favor to localize at  the four corners of the correlation map. The reason is that there is no interference because the system is completely incoherent, which is confirmed by that  the coherent term in Eq.~(\ref{eq:extwocre}) vanishes in the case of  $\eta=0$.  Whereas,  with the increase of the degree of coherence, the coherent term emerges in the two-photon correlation function $\Gamma_{q,r}$, so the $\Gamma_{q,r}$ exhibits the properties of interference.  Due to the existence of the Hanbury Brown-Twiss (HBT) interference, two local maximums  emerge in the off-diagonal  regions of the  correlation matrix  (see Fig.~\ref{fig:tcou} (c) and (d)).  That  implies that the two photons favor to far from each other which is in contrast to the case of Fig.~\ref{fig:tcou} (e) and (f) where except the initial relative phase $\phi$,  the other parameters are the same as those in  Fig.~\ref{fig:tcou} (c) and (d), respectively.  This is reasonable because the coherent term in Eq.~(\ref{eq:twocrebf}) is in proportion to  $\cos{\phi}$ for the periodical waveguide lattice we considered.
Additionally, comparing the values of the maximums in Fig.~\ref{fig:tcou} (c) and (d),
it is easy to find
that the larger the degree of coherence is,
the more distinct the interference effect of the system will be.
\begin{figure}[tbph]
\includegraphics[width=100mm]{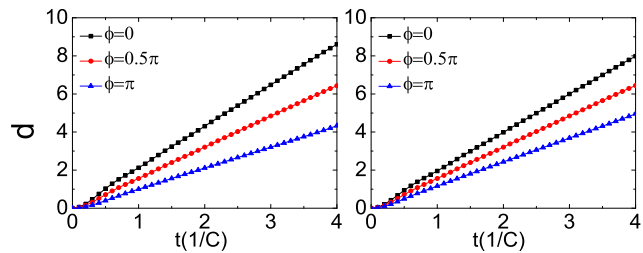}
\caption{(Color online) The time evolution  of the distance between two photons for different initial conditions. The parameters are $\alpha=0.5$, and $\eta=1$ (left panel), $\eta=0.5$ (right panel).}
\label{fig:distanceevo}
\end{figure}
\begin{figure}[tbph]
\includegraphics[width=100mm]{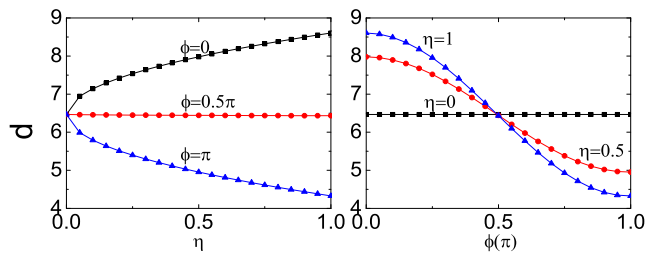}
\caption{(Color online) The dependence of the distance between two photons at time $t=4(1/C)$ on the degree of coherence (left panel) and the initial relative phase (right panel). The parameter is $\alpha=0.5$.}
\label{fig:distance}
\end{figure}

\begin{figure}[tbph]
\includegraphics[width=80mm]{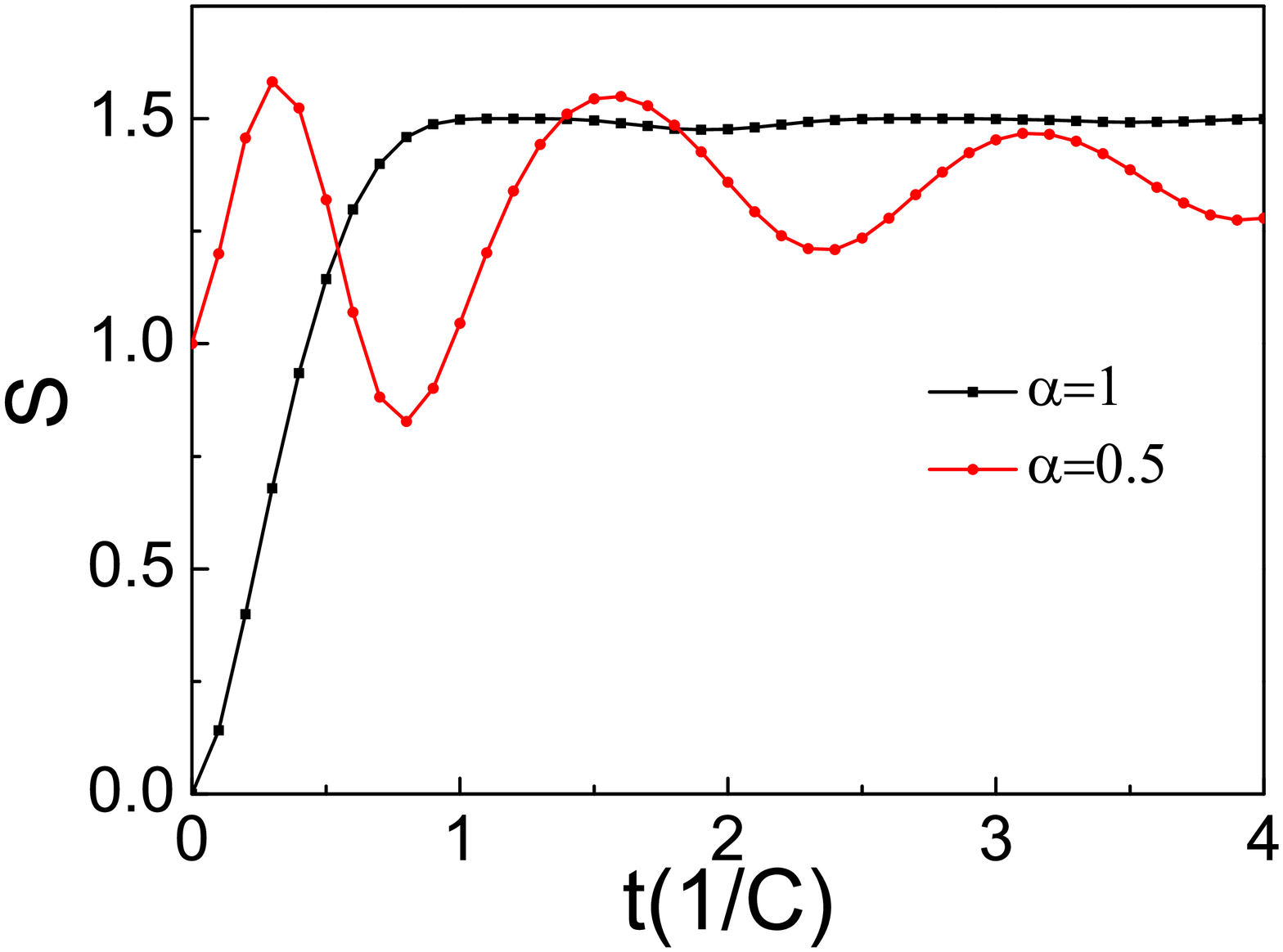}
\caption{(Color online) The time evolution of the von Neumann entropy of the set of  sites on the left  part of the system. The parameters are $\eta=1$, $\phi=0$, and $L=15$.}
\label{fig:von}
\end{figure}

To exhibit the propagation properties of the two photons,
we  also calculate  the average distance  between the two photons,
\begin{eqnarray}\label{eq:distance}
\displaystyle d&=&-2\gamma \cos{\phi}\sum_{q>r}(q-r)J_q(\tau)J_r(\tau)J_{r-1}(\tau)J_{q-1}(\tau)\nonumber\\
&&+\sum_{q>r}(q-r)\Bigl(2\alpha [ J_q(\tau) J_r(\tau) ]^2
   +2(1-\alpha)[ J_{r-1}(\tau) J_{q-1}(\tau) ]^2\Bigr).
   \end{eqnarray}
  Here  the first term is a coherence one that is affected by the degree of coherence of the system    due to $\gamma=\sqrt{\eta-1+4\alpha(1-\alpha)}$.
We plot the time evolution of the distance between the two photons in Fig.~\ref{fig:distanceevo},
we can see that  the distance between two photons is affected not only by the relative phase but also by the degree of coherence.
In Fig.~\ref{fig:distance}, we plot the dependence of the
distance $d$  at time $t=4(1/C)$ on the degree of coherence and the initial relative phase.
From the left panel of this figure, we can see that the distance between two photons becomes larger with the increase of the degree of coherence when $\phi=0$, which is contrast to the case of $\phi=\pi$.
The reason for this phenomenon is  that the HBT interference makes the two photons far from each  when $0\leq\phi<\pi/2$,
which can be confirmed by Fig.~\ref{fig:tcou} (c) and (d) where there are two maximums in the off-diagonal regions.
Therefore, the distance between the two photons becomes larger with  the increase of the degree of coherence due to the fact that the increase of the degree of coherence makes the interference effect more significant.
Whereas, the case of $\pi/2\leq\phi<\pi$ is in contrast to that of $0\leq\phi<\pi/2$ because the two photons favor to stay  together when  $\pi/2\leq\phi<\pi$, which can be confirmed by Fig.~\ref{fig:tcou} (e) and (f).
Note that when $\phi=\pi/2$, the interference term in the two-photon correlation function becomes zeros, so the degree of coherence does not affect the distance between two particles (see the dot-symbol line in the left panel of  Fig.~\ref{fig:distance}).
The right panel of  Fig.~\ref{fig:distance} exhibits that
the relative phase of the system  at the initial time can affect the distance between two photons in the case of $\eta>0$ but such an effect vanishes in the case of $\eta=0$.
That is reasonable because the distance between two photons is in proportion to $\cos{\phi}\sqrt{\eta-1-4\alpha^2+4\alpha}$  which can be found in Eq.~(\ref{eq:distance}).  Additionally, we calculate the von Neumann entropy to show the evolution of the entanglement of the system. We split the system into two halves, $L$ and $R$, in the center of the system,  and build the reduced density matrix $\rho_L$ of the subsystem $L$  at any time~\cite{Schach}.  Then we can calculate the von Neumann entropy of $\rho_L$ as
\begin{equation}
S=-\sum_i\lambda_i\log_2\lambda_i,
\end{equation}
where $\lambda_i$ are the non-zero eigenvalues of the matrix $\rho_L$. In Fig.~\ref{fig:von}, we plot the time evolution of the von Neumann entropy of the left half of the system for different initial conditions.

\section{conclusion}\label{sec:conc}

We  proposed a density matrix formulism to study the properties of  two-particle  quantum walks  where the effect of coherence was introduced naturally. We gave the general analytical expression of the two-particle correlation function which is correct for systems in  both mixed states and  pure states. We suggested  a possible two-photon scheme to exhibit the more fascinating quantum features of two-particle random walks with  mixed initial states. For such a concrete scheme, we calculated  the two-photon correlation and the average distance between the two photons.  The corresponding results manifested  that the propagation of the two photons depends  not only on the initial distribution of the two photons but also on the relative phase and  the degree of  coherence of the system. Such propagation features  of the two photons were  explained  with the help of the analytical expression of the two-particle correlation function we obtained.

The work is supported by the NBRP of China (2014CB921201),  the NSFC (11104244 and 11274272, 11434008),  and by the Fundamental Research Funds for Central Universities.

\end{document}